\documentclass[sigconf,screen,nonacm]{acmart}

\usepackage{multirow}
\usepackage[caption=false,font=footnotesize]{subfig}
\usepackage{array}
\usepackage{xurl}
\usepackage{tabularx}
\usepackage{algorithm}
\usepackage{algpseudocode}
\usepackage[most]{tcolorbox}
\usepackage{enumitem}
\usepackage{balance}

\graphicspath{{figures/}}
\newif\ifdraft
\draftfalse

\ifdraft
  \newcommand{\todo}[1]{{\color{red}\textbf{[TODO: #1]}}}
  \newcommand{\note}[1]{{\color{blue}[Note: #1]}}
  
\else
  \newcommand{\todo}[1]{}
  \newcommand{\note}[1]{}
  
\fi

\newcommand{\Talone}{T^{\text{alone}}}
\newcommand{\Tshared}{T^{\text{shared}}}
\newcommand{\Tqueue}{T^{\text{queue}}}

\newcommand{\sys}{{\fontfamily{lmss}\selectfont Duru}}

\begin{document}

\title{Towards Fair Resource Sharing in Agentic LLM Serving}



\author{Sehyeon Lee}
\affiliation{%
  \institution{University of Illinois Urbana-Champaign}
  \city{Urbana}
  \state{Illinois}
  \country{USA}
}
\email{sehyeon2@illinois.edu}

\author{Yuchen Xia}
\affiliation{%
  \institution{University of Illinois Urbana-Champaign}
  \city{Urbana}
  \state{Illinois}
  \country{USA}
}
\email{yxia28@illinois.edu}

\author{Zhaoyang Hong}
\affiliation{%
  \institution{University of Illinois Urbana-Champaign}
  \city{Urbana}
  \state{Illinois}
  \country{USA}
}
\email{zhong45@illinois.edu}

\author{Jinrui Hu}
\affiliation{%
  \institution{University of Illinois Urbana-Champaign}
  \city{Urbana}
  \state{Illinois}
  \country{USA}
}
\email{jinruih2@illinois.edu}

\author{Anuradha Pandey}
\affiliation{%
  \institution{University of Illinois Urbana-Champaign}
  \city{Urbana}
  \state{Illinois}
  \country{USA}
}
\email{apandey8@illinois.edu}

\author{Evan Zhao}
\affiliation{%
  \institution{University of Illinois Urbana-Champaign}
  \city{Urbana}
  \state{Illinois}
  \country{USA}
}
\email{evanhz2@illinois.edu}

\author{Ankita Nayak}
\affiliation{%
  \institution{Gimlet Labs}
  \city{San Francisco}
  \state{California}
  \country{USA}
}
\email{ankitanayak@gimletlabs.ai}

\author{Nishil Talati}
\affiliation{%
  \institution{University of Illinois Urbana-Champaign}
  \city{Urbana}
  \state{Illinois}
  \country{USA}
}
\email{nishil@illinois.edu}

\renewcommand{\shortauthors}{Lee et al.}

\begin{abstract}

As LLM agents increasingly execute complex, multi-step tasks on shared inference infrastructure, resource contention across the end-to-end LLM serving stack causes severe and uneven slowdowns across agents.
Existing agentic serving systems primarily optimize throughput and latency, with comparatively limited attention to fairness among concurrent agents.
Although fairness has been studied in conventional LLM serving at the token and request levels, these approaches do not fully capture cumulative delays and stateful resource dependencies across successive LLM calls within an agent.
Moreover, interference spans request routing, runtime scheduling, and hardware resources, making agent-level fairness challenging to define and optimize across an end-to-end serving stack.

This paper presents a cross-layer framework to \textit{define} agent-level fairness, \textit{attribute} sources of unfairness, and \textit{optimize} fairness while preserving throughput.
To account for agents with widely varying execution demands, we build theoretical foundation to define fairness using slowdown relative to standalone execution.
Our empirical analysis reveals that routing, runtime scheduling, and hardware contention jointly determine fairness, with performance-oriented policies often leading to prolonged queuing delays.
To optimize fairness, we introduce \sys{} that monitors each agent's slowdown and uses this signal to guide admission and scheduling decisions across the router and runtime.
\sys{} further alleviates starvation through admission exceptions and integrates slowdown-driven autoscaling to adhere to Service-Level Objectives (SLO).
Our evaluation across various benchmarks and hardware platforms shows that \sys{} reduces 99th-percentile normalized slowdown by a geometric mean of 44.4\% (up to 79.5\%) while delivering 91.8--127.4\% of the state-of-the-art system's throughput.
\end{abstract}

\keywords{LLM serving, agentic AI, fairness, scheduling, KV cache}

\maketitle

\section{Introduction}
\label{sec:intro}

An LLM agent places a language model in a loop, repeatedly invoking the model and interacting with an external environment through tools such as Linux terminals, databases, and web APIs to solve complex tasks~\cite{yao2023react}.
Agents are increasingly used today in solving real-world problems in software engineering~\cite{yang2024neurips-sweagent,hong2024metagpt,qian2024acl-chatdev,wang2025iclr-openhands}, deep research~\cite{shao2024naacl-storm,li2025neurips-webthinker,zheng2025emnlp-deepresearcher,openai2025deepresearch}, and computer use~\cite{xie2024neurips-osworld,he2024acl-webvoyager,koh2024acl-visualwebarena,bonatti2025icml-windowsagentarena}.
To complete a single user task, an agent executes successive LLM calls with growing context, typically on inference infrastructure shared with other agents.
Resource contention across the hardware and software layers of this infrastructure can cause some agents to disproportionately delay others, particularly when their computational demands differ.
Consequently, \textit{ensuring fair resource sharing within LLM serving infrastructure is critical for agentic workloads.}

Recent research on agentic LLM serving has primarily focused on improving performance and resource utilization through various techniques including through KV cache reuse, prefix sharing, and program-aware scheduling~\cite{li2026continuum,tiwari2026cachewise,pan2025kvflow,kang2026thunderagent,luo2026agentix}.
While these optimizations improve aggregate efficiency, they can also favor agents with reusable contexts, potentially at the expense of others.
Fairness has also been studied in LLM serving: VTC~\cite{sheng2024vtc} balances token-level service across clients, DLPM~\cite{cao2025dlpm} incorporates prefix locality into fair scheduling, and FairInference~\cite{bali2026fairinference} targets per-token latency isolation.
More recently, SAGA~\cite{guo2026saga} considered agent-workflow fairness through scheduling based on predicted remaining work and deadlines.
However, whether these mechanisms adequately address fairness over an agent's entire execution remains unclear.
This raises a fundamental question: \textit{Do existing LLM serving fairness mechanisms work out of the box for agentic workloads?}

Extending fairness from individual LLM requests to complete agent executions presents three fundamental challenges.
First, agents are \textit{stateful}: delays accumulate across dependent LLM requests~\cite{liu2026azure,zhu2026tracelab}, so request-level fairness does not necessarily translate into fairness in end-to-end task completion.
Second, agents are \textit{heterogeneous}: the computational cost of processing tokens varies with context length, inference phase, and resource contention~\cite{yuan2026agentic,patel2024splitwise,chang2026agentsysbench}, making token counts insufficient to capture the actual service received by each agent.
Third, agents are \textit{unpredictable}: their future execution depends on intermediate decisions, including failures that trigger additional reasoning and retries~\cite{liu2026azure,zhu2026tracelab}, and the strength of the model-harness combination, making remaining work difficult to estimate with high accuracy, as required by prediction-based approaches such as SAGA~\cite{guo2026saga}.
\textit{Addressing these challenges requires measuring cumulative agent-level slowdown, accounting for heterogeneous computational costs, and correcting unfairness online without predicting future execution.}

\textbf{In this paper}, we present a \textit{principled framework for understanding and optimizing fairness in agentic LLM serving} that captures cumulative delays and heterogeneous computational costs without requiring predictions of future agent execution.
The goals of our framework are threefold.
First, we seek to establish a theoretical foundation for fairness in single- and multi-agent workflows that distinguishes unfair resource allocation from differences in task complexity and overall system load.
Second, we aim to provide empirical evidence of how hardware contention, runtime scheduling, and routing decisions interact to create unfairness across the end-to-end serving stack.
Third, we aim to translate these insights into a coordinated, fairness-aware serving system that alleviates starvation while limiting throughput degradation.

We show how to establish slowdown-based fairness from shared hardware~\cite{mutlu2007stfm,subramanian2015asm} to agentic workloads, capturing cumulative delays across dependent LLM calls and multi-agent workflows.
By comparing observed execution against estimated standalone performance and ideal processor sharing, our metric distinguishes unfair resource allocation from intrinsic task complexity and overall system load.
We estimate this metric online from observed execution progress, without predicting future agent behavior.
Our empirical analysis across the hardware, runtime, and routing layers reveals a fundamental tension: \textit{optimizations that improve serving efficiency can inadvertently worsen fairness.}
Long contexts amplify interference, KV cache retention can starve waiting agents, and admission control can shift delays from GPU execution to router queues.

This work's objective is to minimize 99th percentile (p99) agent-level unfairness while limiting throughput degradation.
To this end, we design \sys{}, a fairness-aware serving system that coordinates scheduling and admission control across the router and runtime.
Using online estimates of agent-level normalized slowdown, \sys{} prioritizes agents that have fallen behind their fair share, rather than favoring requests solely for their immediate execution efficiency.
To balance fairness and throughput, \sys{} combines slowdown-aware scheduling with capacity-aware resource allocation, preserving KV cache reuse while introducing targeted admission exceptions to alleviate starvation.
However, fairness optimization alone cannot resolve performance degradation caused by insufficient capacity.
To address this, \sys{} incorporates SLO-driven autoscaling that adjusts GPU capacity based on agent-level slowdown under varying workloads.
Collectively, these mechanisms improve fairness and SLO attainment while maintaining high throughput.

We evaluate \sys{} using three LLMs (Qwen3-Coder-30B-A3B, Llama-3.3-70B, and DeepSeek-V4-Flash) and three real-world agentic workload traces: Azure~\cite{liu2026azure}, TraceLab~\cite{zhu2026tracelab}, and AgentX~\cite{semianalysis2026agentx}.
We use both NVIDIA and AMD GPUs to demonstrate that \sys{} is vendor-agnostic.
We compare against state-of-the-art production frameworks, including NVIDIA Dynamo~\cite{dynamo} and llm-d~\cite{llmd}, as well as recent agent-aware scheduling and KV cache management techniques~\cite{li2026continuum,tiwari2026cachewise}.
Our evaluation shows that \sys{} improves p99 fairness by a geometric mean of 44.4\% (up to 79.5\%) while maintaining 91.8--127.4\% of baseline throughput.
In a 48-hour replay of the production trace~\cite{liu2026azure}, \sys{} achieves 97.4\% agent-level SLO attainment with comparable throughput while using fewer GPUs than NVIDIA Dynamo.
Our key contributions are as follows.


\begin{itemize}[leftmargin=*]
    \item \textbf{Definition.} We formulate slowdown-based fairness for single and multi-agent workflows, capturing unfairness relative to ideal resource sharing without predicting future execution.
    \item \textbf{Attribution.} We reveal how performance-oriented optimizations across hardware, runtime, and routing layers can amplify unfairness or shift it between layers, leading to agent starvation.
    \item \textbf{Optimization.} We design \sys{} to reduce p99 unfairness via coordinated, slowdown-aware scheduling and admission control with limiting throughput loss, and SLO-driven autoscaling.
\end{itemize}
\section{Background}
\label{sec:background}

\begin{figure}[t]
  \centering
  \includegraphics[width=\columnwidth]{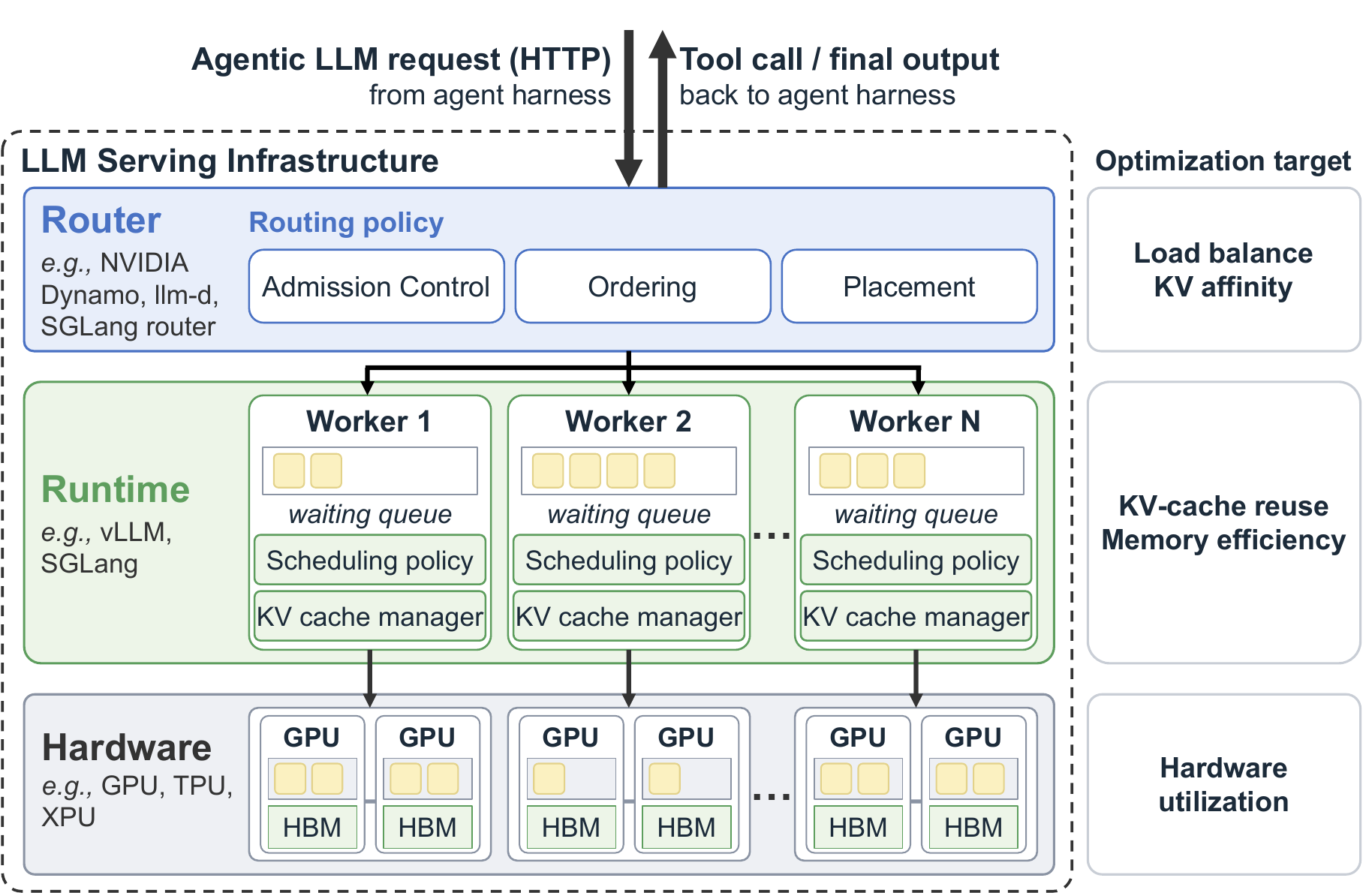}
  \vspace{-0.7cm}
  \caption{An end-to-end agentic LLM serving stack.
  }
  \vspace{-0.5cm}
  \label{fig:serving-arch}
\end{figure}

\subsection{LLM Agent}
\label{sec:llm-agent}
An LLM agent places a language model in a loop to interact with an environment, such as a Linux terminal, database, or web browser, using tool outputs to guide LLM calls.
The \emph{agent harness}~\cite{yang2024neurips-sweagent,minisweagent2025,deepseek-harness2026,anthropic2025claudecode,openai2025codex} orchestrates LLM calls and tool execution, passing results between the model and its environment.
In a single-agent system~\cite{yao2023react}, one agent orchestrates this loop, \textit{e.g.,} a coding agent reads files, edits code, and executes tests.
Multi-agent systems~\cite{hong2024metagpt,wu2024autogen,qian2024acl-chatdev} instead decompose tasks into interdependent subtasks assigned to specialized agents that exchange results and may execute concurrently.

In both settings, an agentic loop alternates between LLM calls and tool execution.
As an agent executes successive LLM calls, its context grows with instructions, LLM outputs, and tool results, much of which forms a reusable prefix~\cite{liu2026azure,yuan2026agentic}.
Reusing the attention keys and values (KV cache) computed for this prefix avoids redundant prefill computation in subsequent calls~\cite{kwon2023vllm,gao2024cachedattention}.
This paper focuses on understanding and optimizing fairness across the \emph{end-to-end LLM serving stack}.

\subsection{LLM Serving Stack for Agentic Inference}
\label{sec:serving-stack}


Figure~\ref{fig:serving-arch} illustrates the end-to-end LLM serving stack, comprising three layers: router, runtime, and hardware.
For each LLM call, the agent harness sends an HTTP request and receives generated tokens.
After a tool call, the harness may return with updated context. We describe the three serving layers below.

\noindent\textbf{Router layer.}
The routing layer is typically the first scheduling stage encountered by an incoming HTTP request, dispatching it to one or more inference workers.
Systems such as NVIDIA Dynamo~\cite{dynamo}, llm-d~\cite{llmd}, and SGLang Model Gateway~\cite{sglang} support distributed request routing.
A \textit{primary objective} is to minimize latency while balancing worker utilization against KV cache locality.
Load-based routing~\cite{srivatsa2025preble, dynamo} distributes requests to avoid queuing but may scatter successive calls from the same agent across workers, losing KV cache reuse opportunities.
Conversely, cache-aware routing preserves reusable context prefixes at the risk of overloading cache-rich workers.
Modern routers~\cite{dynamo,llmd} combine cache affinity with worker load and may perform admission control or select separate prefill and decode workers.
Agent-aware policies~\cite{wang2026smetric,kang2026thunderagent} further exploit session- and program-level information to coordinate placement across successive LLM calls.

\noindent\textbf{Runtime layer.}
Within each inference worker, the runtime schedules queued requests and manages KV caches.
Inference engines such as vLLM~\cite{kwon2023vllm} and SGLang~\cite{zheng2024sglang} employ first-come, first-served (FCFS) scheduling with continuous batching.
A \textit{primary objective} is to maximize throughput while minimizing latency through efficient GPU computation and KV cache utilization.
For agentic workloads, retaining KV caches during tool execution avoids redundant prefill but occupies GPU memory while agents are idle.
Agent-aware systems~\cite{luo2026agentix,li2026continuum,tiwari2026cachewise,pan2025kvflow,bian2026tokencake,pan2026scalesim,liu2026askthetool} coordinate scheduling and KV-cache management using execution dependencies and tool calls.

\noindent\textbf{Hardware layer.}
LLM serving runs on hardware accelerators such as GPUs~\cite{choquette2023hopper,jarmusch2026blackwell}, TPUs~\cite{jouppi2017tpu,jouppi2023tpuv4}, and specialized AI accelerators~\cite{srivastava2025corsair,nair2026raptor,lie2024cerebras}.
Each inference worker runs a model on one or more accelerators, using model parallelism and high-bandwidth interconnects when spanning multiple devices~\cite{narayanan2021megatron,rajbhandari2022deepspeedmoe,nvidia_nvlink}.
Within a worker, continuously batched requests contend for compute and memory bandwidth, while agents' growing contexts and retained KV caches place additional pressure on limited GPU memory (HBM).
To address these limitations, systems can offload KV caches to host DRAM or SSD~\cite{qin2025mooncake,liu2025lmcache}, or disaggregate prefill and decode across separate workers to reduce interference~\cite{zhong2024distserve,patel2024splitwise}.


\begin{table}[t]
\centering
\footnotesize
\caption{LLM serving tiers and resource sharing.}
\vspace{-0.35cm}
\label{tab:serving-tiers}
\setlength{\tabcolsep}{3pt}
\renewcommand{\arraystretch}{1.0}
\begin{tabularx}{\columnwidth}{@{}l
    >{\raggedright\arraybackslash}X
    >{\raggedright\arraybackslash}p{2.25cm}@{}}
\toprule
\textbf{Tier} & \textbf{Description} & \textbf{Examples} \\
\midrule
Dedicated
  & Isolated, single-tenant GPU inference.
  & Together AI, Nebius~\cite{togetherDedicated,nebiusDedicated} \\
Reserved
  & Guaranteed throughput; reserved capacity.
  & Azure, Google~\cite{azureProvisioned,gcpThroughput} \\
Priority
  & Preferential service with lower latency.
  & Azure, Google, AWS~\cite{azurePriority,gcpPriority,awsTiers} \\
Standard
  & Shared capacity; throughput varies with load.
  & Google, AWS~\cite{gcpPriority,awsTiers} \\
Flex
  & Discounted; higher latency or throttling.
  & Google, AWS~\cite{gcpFlex,awsTiers} \\
Batch
  & Discounted asynchronous bulk inference.
  & Together AI, Alibaba~\cite{togetherBatch,alibabaBatch} \\
\bottomrule
\end{tabularx}
\end{table}

\subsection{Real-World LLM Serving}
\label{sec:service-tiers}


LLM serving operates under different degrees of resource sharing and performance isolation.
Our survey of LLM providers (Table~\ref{tab:serving-tiers}) identifies service tiers that trade off cost, latency, and isolation.
At one extreme, dedicated endpoints provide exclusive GPUs to individual tenants~\cite{togetherDedicated,nebiusDedicated}, while guaranteed-throughput offerings reserve capacity~\cite{azureProvisioned,gcpThroughput}.
Priority tiers offer preferential processing and predictable latency on shared infrastructure~\cite{azurePriority,gcpPriority,awsTiers}.
In contrast, standard, flex, and batch tiers give providers greater scheduling flexibility, with flex and batch offerings trading higher latency or asynchronous completion for lower cost~\cite{gcpFlex,awsTiers,togetherBatch,alibabaBatch}.

While service tiers differ in their performance guarantees, shared infrastructure exposes requests to interference from other tenants.
Serving engines such as vLLM~\cite{kwon2023vllm} and TensorRT-LLM~\cite{nvidia2023tensorrtllm} employ continuous batching to improve GPU utilization, but introduce co-batch interference~\cite{kwon2023vllm,nvidia_trtllm_scheduling}.
Latency-sensitive tiers may limit such interference by reserving capacity or reducing worker load~\cite{sun2024llumnix}, whereas best-effort tiers can favor higher utilization through more aggressive batching.
Thus, \emph{performance guarantees determine resource sharing and acceptable interference across tiers}.




\subsection{Fairness in Resource Scheduling}
\label{sec:fairness}
Fairness has been extensively studied across shared computing resources.
In DRAM and LLCs, fairness mechanisms~\cite{mutlu2007stfm,mutlu2008parbs,ebrahimi2010fst,lu2015dynamicassym} aim to minimize interference-induced slowdowns while preserving throughput.
Operating system schedulers~\cite{linux2007cfs,linux2024eevdf} approximate equal or weighted processor sharing through virtual-runtime and deadline-based scheduling.
Similarly, network schedulers approximate generalized processor sharing through fair queuing, with flow slowdown used to characterize interference~\cite{parekh1993gps,demers1989wfq,shreedhar1995drr,montazeri2018homa}.
In LLM serving, VTC~\cite{sheng2024vtc} tracks per-client service using weighted input and output tokens, while DLPM~\cite{cao2025dlpm} additionally considers KV cache locality to balance fairness and efficiency.
FairInference~\cite{bali2026fairinference} extends beyond service fairness to provide token-level latency isolation by accounting for GPU computation and KV cache interference.

\noindent\textbf{Fairness for agentic workloads.}
Unlike single-turn inference, agentic workloads involve dependent LLM calls with varying context lengths and concurrent subagents~\cite{kim2026sciencescalingagentsystems,liu2026azure,zhu2026tracelab,bian2026tokencake}.
Interference and queuing delays accumulate across calls, so token- or request-level fairness does not necessarily ensure agent-level fairness.
Recent agent-aware systems exploit program structure to improve scheduling~\cite{luo2026agentix,kang2026thunderagent,wang2026smetric}, while SAGA~\cite{guo2026saga} addresses workflow-level fairness.
However, existing approaches do not jointly measure and optimize agent-level slowdown from routing, runtime scheduling, KV cache management, and hardware contention, motivating fairness coordination across the \emph{entire LLM serving stack}.

\section{Defining Fairness for Agentic LLM Serving}
\label{sec:metric}

An absolute latency deadline is ill-suited for assessing fairness across agent sessions, whose execution time depends on task complexity, the model, and the agent harness~\cite{kim2026sciencescalingagentsystems,yang2024neurips-sweagent,yuan2026agentic,nixon2026requests}.
Agent workloads range from a few calls to hundreds, spanning seconds to hours~\cite{yuan2026agentic,liu2026azure,nixon2026requests,zheng2026agentserving}.
A fixed latency target would therefore conflate intrinsic task complexity with delays imposed by resource sharing.
Instead, we employ \emph{slowdown}, an established fairness metric in shared-resource systems~\cite{mutlu2007stfm,ebrahimi2010fst}, comparing execution time under sharing (including queuing) against standalone execution time.
Applying this principle to agentic workloads requires accounting for cumulative delays across dependent LLM calls and fair-share entitlements among concurrent sub-agents.
To this end, we establish a unified theoretical formulation for single- and multi-agent workflows relative to ideal processor sharing, enabling online fairness estimation and cross-layer optimization.

\begin{definition}[Single-agent slowdown]
\label{def:slowdown}
For an agent session $i$, let $\Talone_i$ denote its execution time
when served alone, $\Tshared_i$ its execution time under sharing, and
$\Tqueue_i$ its cumulative waiting time before and between execution
intervals. The \emph{slowdown} of session $i$ is
\begin{equation}
    S_i = \frac{\Tshared_i + \Tqueue_i}{\Talone_i} \geq 1.
    \label{eq:slowdown}
\end{equation}
\end{definition}

We define slowdown in the \textit{time domain} because GPU requests contend for multiple resources, including compute, memory bandwidth, memory capacity, and interconnect bandwidth.
Since these resources have different units, quantifying their combined interference is challenging.
Time provides a common measure: contention for any resource manifests as increased latency, whether through queuing delays or slower execution.
Thus, slowdown captures the impact of resource sharing regardless of the source of contention.

For multi-agent tasks, we aggregate the root agent $r$ and spawned agents’ slowdowns, weighted by their shares of standalone work.
\begin{definition}[Multi-agent slowdown]
\label{def:task-slowdown}

For a task initiated by root agent $r$, let $\mathcal{A}(r)$ denote the set of $r$ and all its spawned agents. 
For each $j\in\mathcal{A}(r)$, define
\begin{equation*}
    w_j = \frac{\Talone_j}{\sum_{k\in\mathcal{A}(r)}\Talone_k}.
\end{equation*}
The \emph{task slowdown} of $r$ is
\begin{equation}
\begin{aligned}
    S_r
    &= \sum_{j\in\mathcal{A}(r)} w_j S_j
    &= \frac{\sum_{j\in\mathcal{A}(r)}
                  (\Tshared_j+\Tqueue_j)}
            {\sum_{j\in\mathcal{A}(r)}\Talone_j}.
\end{aligned}
\label{eq:multi-agent}
\end{equation}
\end{definition}

The standalone-time weights cancel in the second expression, so task slowdown is the ratio of aggregate inference time under sharing to aggregate standalone inference time. 

\noindent\textbf{Normalized slowdown.}
\textit{Absolute slowdown conflates unfair scheduling with heavy load}: a job can slow down despite receiving its fair share. 
We therefore normalize against \emph{ideal processor sharing} (PS)~\cite{parekh1993gps}, which divides capacity equally among active jobs.

\begin{definition}[Normalized slowdown, $U_i$]
\label{def:ps-fairness}
Let $G(t)$ be the number of available workers, $N(t)$ the number of root agents with at least one outstanding (waiting or running) request, and $k_r(t)$ the number of agents in root agent $r$'s multi-agent session with at least one outstanding request, all at time $t$.
Let $\mathcal{I}_j$ be the times agent $j$ has a request outstanding, excluding tool-execution gaps, so that $|\mathcal{I}_j|=\Tqueue_j+\Tshared_j$.
In the PS reference, each multi-agent job receives an equal share $G/N(t)$ of the workers, split equally among its $k_r(t)$ outstanding agents, and no agent exceeds one worker, so agent $j$ associated with root agent $r$ receives a fraction
\begin{equation*}
    \rho_j(t) = \min\!\left\{1,\ \frac{G(t)/N(t)}{k_r(t)}\right\}
\end{equation*}
of its standalone processing rate.
We define the \emph{PS-relative slowdown} of agent $j$ and the multi-agent session associated with root agent $r$ (with $w_j$ as in Definition~\ref{def:task-slowdown}) as
\begin{equation}
    U_j = \frac{1}{\Talone_j}\int_{\mathcal{I}_j}\rho_j(t)\,dt,
    \qquad
    U_r = \sum_{j\in\mathcal{A}(r)} w_j U_j .
\label{eq:ps_fairness}
\end{equation}
\end{definition}

Thus, $U_i$ is the ratio of the service job $i$ was entitled to under PS to the standalone-equivalent work it completed.


\noindent \textbf{Estimating \texorpdfstring{$\Talone$}{Talone}.}
\label{sec:talone}
$\Talone_i$ cannot be directly observed at runtime under resource sharing.
We estimate it using offline measurements of isolated prefill and decode latencies under the same model, hardware, and serving configuration.
At runtime, we interpolate these measurements to estimate each iteration's standalone latency and sum them to obtain $\Talone_i$.
Interpolation is effective because the latency of memory-bound kernels grows approximately linearly with the amount of KV cache data accessed, particularly during decoding and prefill with long cached prefixes~\cite{yuan2026agentic,liu2026azure}.


\textbf{Decode.}
 We model isolated single-iteration latency by KV-cache size and linearly interpolate between measured points.

\textbf{Prefill.}
We measure prefill latency across new token counts and cached prefix sizes, constructing a two-dimensional grid.
We estimate latency through bilinear interpolation between grid points.

\textbf{Practicality.}
Calibration is a one-time cost per deployment configuration, as the resulting latency tables are independent of workloads.
Estimating $\Talone_i$ incrementally at each iteration also eliminates the need to know a request's output length in advance.


\textbf{Validation.}
We validate $\Talone$ by comparing estimated and measured inference times on GPUs using agent traces from Azure's GitHub Copilot~\cite{liu2026azure}, TraceLab~\cite{zhu2026tracelab}, and AgentX~\cite{semianalysis2026agentx}.
On Qwen3-Coder-30B-A3B (RTX PRO 6000), 534 Azure and TraceLab sessions yield mean absolute relative errors of 1.9\% and 1.8\%, respectively, with maximum errors of 4.1\% and 4.2\%.
On Llama-3.3-70B (2$\times$H100, TP=2), 200 sessions totaling 1.95M output tokens yield mean errors of 0.2\%, 0.3\%, and 0.1\% for Azure, TraceLab, and AgentX, respectively, with maximum errors of 1\%, 2.6\%, and 0.3\%.

\begin{definition}[Live slowdown]
\label{def:live-slowdown}
Let $K_j(t)$ be the number of LLM iterations agent $j$ has completed by time $t$, and $\hat{t}_{j,k}$ the estimated standalone latency of its $k$th iteration, interpolated from its number of new tokens and KV cache size.
Its estimated completed work is $\hat{W}_j(t) = \sum_{k=1}^{K_j(t)} \hat{t}_{j,k}$.
When $\hat{W}_r(t) = \sum_{j\in\mathcal{A}(r)} \hat{W}_j(t)>0$, the \emph{live slowdown} $\hat{S}_r(t)$ and \emph{live normalized slowdown} $\hat{U}_r(t)$ of task $r$ are
\begin{equation}
    \frac{1}{\hat{W}_r(t)} \sum_{j\in\mathcal{A}(r)} \int_{\mathcal{I}_j \cap [0,t]} \rho_j(t')\,dt',
    \label{eq:live-slowdown}
\end{equation}
with $\rho_j \equiv 1$ for $\hat{S}_r(t)$ and $\rho_j$ as in Definition~\ref{def:ps-fairness} for $\hat{U}_r(t)$.
\end{definition}

This estimate updates from observed waiting and completed work, without predicting future requests or remaining time.

\textbf{Intuition.}
A normalized slowdown of $U_i=1$ indicates that an agent received its fair share under processor sharing.
A value $U_i>1$ indicates the agent fell behind its fair share, whereas $U_i<1$ indicates it received more.
These values identify disproportionately delayed agents, enabling fairness-aware system design.

\section{Attributing Sources of Unfairness} \label{sec:attributing_unfrirness}

We posit that agent-level unfairness arises from interacting effects across the \emph{hardware}, \emph{runtime}, and \emph{router} layers of the serving stack (Figure~\ref{fig:serving-arch}).
For a given workload, model, and hardware, the cumulative slowdown of agent session $i$ depends on
\begin{equation}
  S_i = f\big(\mathbf{c},\; b,\; \boldsymbol{\ell},\; \Lambda\big),
  \label{eq:slowdown-factors}
\end{equation}
where, for the $n$ outstanding requests, $\mathbf{c}=(c_1,\dots,c_n)$ denotes their context lengths, $b \le n$ the number running, $\boldsymbol{\ell}=(\ell_1,\dots,\ell_n)$ their KV cache locations (GPU, CPU, or discarded), and $\Lambda$ the prefill--decode (PD) placement.
The \emph{runtime} selects which $b$ requests run, leaving the rest queued, and shapes $\boldsymbol{\ell}$ via cache retention and eviction.
The \emph{router} shapes $\boldsymbol{\ell}$ by steering an agent's calls toward or away from the worker holding its context, and sets $\Lambda$: whether and how prefill and decode are disaggregated.
At the \emph{hardware} layer, running requests contend for GPU compute, memory bandwidth, and capacity, so their number and context lengths set how much they slow one another.
All four factors evolve throughout an agent's execution.

Intuitively, the context lengths $\mathbf{c}$ determine the computation and memory bandwidth required for each inference iteration.
Long-context requests increase KV cache accesses, slowing other requests in the same batch, while their larger memory footprint can also limit concurrency.
The number of running requests $b$ determines the demand for GPU resources.
Larger batches increase per-iteration computation and, for MoE models, expert-weight traffic, while requests that cannot be scheduled must wait, increasing queuing delay $T_{\text{queue}}$.
The KV cache locality $\boldsymbol{\ell}$ determines whether previously computed context can be reused or must be reloaded or recomputed.
Runtime cache management and router placement influence this locality, trading reduced computation against GPU memory pressure and potential starvation of waiting agents.
Finally, the PD placement $\Lambda$ determines how prefill and decode workloads share GPU resources.
Co-locating the two phases introduces interference that slows token generation, while disaggregation reduces this interference but may increase queuing at either phase.
Next, we quantify these effects at each serving layer.

\begin{figure}[t]
  \centering
   \includegraphics[
        width=\columnwidth, trim=10pt 0pt 0pt 0pt, clip]{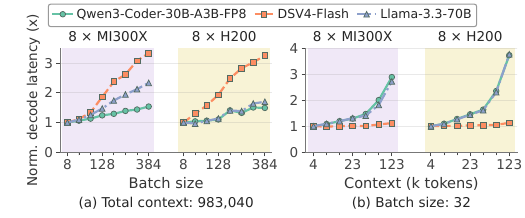}
\vspace{-0.75cm}
\caption{Normalized decode latency on 8 MI300X and H200 GPUs (TP=8; EP=8 for MoE): (a)~varying batch size at 983,040 total tokens; (b)~varying context at batch size 32. Decode latency depends on batch size, context length, and model.}  
\Description{}
\label{fig:batch-ctx}
\end{figure}

\begin{figure}[t]
  \centering
  \includegraphics[
        width=\columnwidth, trim=4pt 0pt 0pt 0pt, clip]{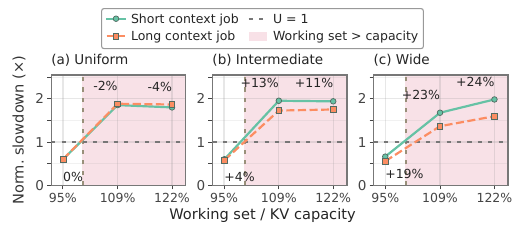}
\vspace{-0.75cm}
\caption{Fairness versus KV cache pressure under FCFS with LMCache on H100 GPUs, for
  (a) uniform (96K/96K), (b) intermediate (64K/128K), and (c) wide (16K/176K)
  contexts. Percentages show the short-context group's normalized slowdown
  relative to the long-context group's.}
  \vspace{-0.5cm}
  \Description{}
  \label{fig:hetero-ctx}
\end{figure}

\subsection{Hardware Contention and Unfairness}
\label{sec:hardware-layer}
Finite GPU compute, HBM capacity, memory bandwidth, and interconnect bandwidth create contention among concurrently executing requests.
Because co-batched requests advance in lockstep, long-context requests slow the entire batch by increasing compute and memory bandwidth demand for each prefill and decode phase.
These delays disproportionately slow short-context requests, whose standalone execution time is smaller.
Larger batches further increase resource demand; for MoE models, expert-weight traffic and inter-GPU communication.
Consequently, identical requests can experience different slowdowns depending on the resource demands of their co-runners and the load on their assigned workers.

We examine batch size and context length effects on decode latency using equal-context requests within each batch.
We focus on decode because agentic workloads exhibit substantial KV cache reuse across LLM calls~\cite{yuan2026agentic,liu2026azure}, making decode the dominant inference phase when contexts remain resident~\cite{yuan2026agentic}.
We measure Qwen3-Coder-30B-A3B (Qwen), DeepSeek-V4-Flash (DeepSeek), and Llama-3.3-70B (Llama) on 8 AMD MI300X and 8 NVIDIA H200 GPUs with TP=8 and EP=8 for the MoE models (Figure~\ref{fig:batch-ctx}).

In Figure~\ref{fig:batch-ctx}(a), we hold aggregate context fixed at 983,040 tokens while increasing batch size from 8 to 384, reducing the context per request.
This isolates the effect of processing more concurrent requests without increasing aggregate KV cache demand.
DeepSeek's decode latency increases by 3.3$\times$, versus 1.5$\times$ for Qwen and 1.7--2.3$\times$ for Llama.
DeepSeek's increase stems from MoE communication: on MI300X, its all-to-all time rises from 12.9 to 74.7 ms, versus 5.4 to 7.2 ms for Qwen.
Conversely, Figure~\ref{fig:batch-ctx}(b) fixes batch size at 32 while increasing per-request context from 3,840 to 122,880 tokens, isolating the effect of growing KV cache demand at fixed concurrency.
Qwen and Llama latency rises by 2.7--3.8$\times$, while DeepSeek increases by only 1.1$\times$ as its compressed attention~\cite{deepseek2026v4} reduces KV cache accesses.
These results show that the \textit{per-token cost varies} with batch size, context length, and model architecture.

Next, we examine how mixing requests with different context lengths leads to agent-level unfairness under hardware contention.
Figure~\ref{fig:hetero-ctx} compares three workloads with heterogeneous prompt lengths. 
Each workload is evenly divided into two job groups, with prompt lengths of 96K/96K tokens for uniform, 64K/128K for intermediate, and 16K/176K for wide contexts.
All three workloads have the same average prompt length of 96K tokens.
We use Qwen with TP=2, FCFS scheduling, and LMCache~\cite{liu2025lmcache}, with each job generating 256 tokens per request and waiting 10 seconds between requests.
Increasing concurrency from 14 to 18 raises working-set demand from 95\% to 122\% of GPU KV cache capacity.

Under uniform contexts, both groups experience similar normalized slowdown ($U$).
However, the short-context group's $U$ exceeds the long-context group's by 4--13\% under intermediate and 19--24\% under wide contexts, with the gap widening under memory pressure.
This disparity arises from two mechanisms.
First, without memory pressure, long-context requests increase shared decode iteration time, imposing a larger slowdown on short-context jobs with smaller $\Talone$.
Second, as memory demand exceeds capacity, runtime engine introduces queuing delays.
Because $U$ normalizes delay by standalone execution time, the same waiting time disproportionately penalizes short-context jobs, amplifying unfairness.

\begin{center}
\begin{tcolorbox}[
    width=\columnwidth,
    boxsep=1pt,
    left=2pt,
    right=2pt,
    top=4pt,
    bottom=4pt,
    halign=center,
    fontupper=\small
]
\centering
 \emph{\textbf{Takeaway \#1:} Per-token decode cost varies with batch size and context length, with sensitivity differing across models. The mix of co-located requests can disproportionately amplify some requests' latency, leading to unfairness.}
\end{tcolorbox}
\end{center}

\begin{figure}[t]
  \centering
  \includegraphics[width=\columnwidth]{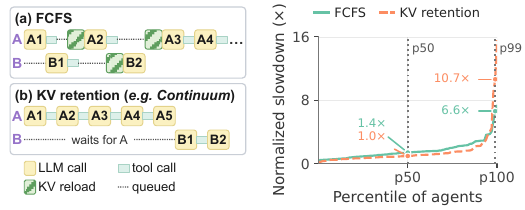}
  \vspace{-0.75cm}
\caption{Left: Agent timelines under (a)~FCFS and (b)~KV-cache retention: retaining A’s cache avoids reloads but prolongs B’s waiting. Right: Normalized slowdown for 100 Azure agents at concurrency 18, using Qwen on an RTX PRO 6000.}

\label{fig:agent-scheduling}
\Description{}
\end{figure}

\begin{figure}[t]
  \centering
  \includegraphics[width=\columnwidth]{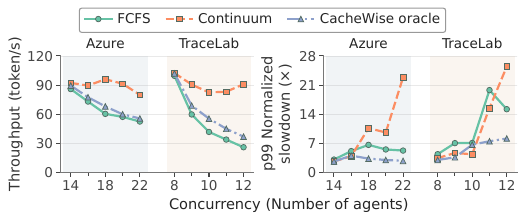}
  \vspace{-0.75cm}
    \caption{Throughput and p99 normalized slowdown of FCFS, Continuum~\cite{li2026continuum}, and the CacheWise oracle~\cite{tiwari2026cachewise} on Azure and TraceLab (one RTX PRO 6000). Concurrency is set per trace to yield comparable memory pressure. Continuum improves throughput but worsens fairness at high concurrency.}
    \vspace{-0.5cm}

\label{fig:agent-azure-scheduler}
\Description{}
\end{figure}

\subsection{Unfairness From Runtime Scheduling}
\label{sec:runtime-layer}
Inference runtimes such as vLLM~\cite{kwon2023vllm} and SGLang~\cite{zheng2024sglang} schedule requests for execution and manage their KV caches in GPU memory.
When the working set exceeds GPU HBM capacity, the KV caches of some requests must be evicted, requiring recomputation or reloading from secondary storage (e.g., CPU DRAM) when those requests resume~\cite{qin2025mooncake,liu2025lmcache,alvarez2026kvcr}.
For agentic workloads, retaining KV cache across tool calls avoids redundant prefill but reserves HBM for idle agents.
Figure~\ref{fig:agent-scheduling} illustrates this trade-off; assuming each agent's KV cache occupies nearly all available GPU memory, retaining agent A's cache can leave agent B waiting across multiple turns, whereas FCFS interleaves their requests at the cost of repeated cache reloads or recomputation.
Empirically, KV cache retention reduces median normalized slowdown from 1.4$\times$ in FCFS to 1.0$\times$ but increases p99 slowdown from 6.6$\times$ in FCFS to 10.7$\times$, \textit{demonstrating how improving common-case performance can worsen p99 fairness.}

We further examine two state-of-the-art cache-aware runtime policies: Continuum~\cite{li2026continuum} and CacheWise~\cite{tiwari2026cachewise}.
Continuum retains KV cache across tool calls using time-to-live estimates, favoring reuse by returning agents but potentially delaying new admissions.
CacheWise instead prioritizes requests with low additional KV cache demand and evicts blocks with the most distant future reuse; we evaluate its oracle variant.
Both optimize cache efficiency, but neither explicitly accounts for cumulative agent-level slowdown.

Figure~\ref{fig:agent-azure-scheduler} compares these policies with FCFS over 100-agent concurrency sweeps.
As concurrency increases, Continuum reaches 1.5$\times$ and 3.5$\times$ FCFS's throughput on Azure and TraceLab, respectively.
However, its p99 $U$ rises from 2.7 to 22.9 on Azure and from 3.5 to 25.5 on TraceLab.
Avoiding repeated prefill raises throughput, but retaining some agents' KV cache can starve others as memory pressure grows.
CacheWise achieves lower throughput than Continuum but lower p99 $U$.
Because $U$ normalizes delay by standalone time, the same delay inflates a short job's slowdown more (\S\ref{sec:hardware-layer}); by serving inexpensive requests first, CacheWise avoids many such slowdowns.
However, the benefit is partial, as per-request cost does not reflect agent-level cost.
Thus, runtime efficiency does not necessarily translate into fair progress across agents.

\begin{center}
\begin{tcolorbox}[
    width=\columnwidth,
    boxsep=1pt,
    left=2pt,
    right=2pt,
    top=4pt,
    bottom=4pt,
    halign=center,
    fontupper=\small
]
\centering
\emph{\textbf{Takeaway \#2:} KV cache retention that improves throughput can also cause agent starvation, worsening p99 fairness.}
\end{tcolorbox}
\end{center}

\subsection{Unfairness Due to Routing Decisions}
\label{sec:router-layer}

The router determines request placement and admission, trading off KV cache locality against worker load.
We evaluate three policies: NVIDIA Dynamo's KV-aware policy~\cite{dynamo}, which prioritizes cache affinity; llm-d::VTC~\cite{llmd-wfs}, which adds admission control based on VTC's token-level fairness~\cite{sheng2024vtc}; and NVIDIA Dynamo's ThunderAgent~\cite{kang2026thunderagent}, which uses program-level working-set limits to control admission.
We replay TraceLab traces~\cite{zhu2026tracelab} on 4$\times$ NVIDIA RTX PRO 6000 Blackwell GPUs, each serving Qwen, sweeping concurrency from 32 to 48 agents in steps of 4 until 200 agents complete per configuration.
Across this range, aggregate KV cache demand increases from approximately 68\% to 96--102\% of capacity.

\begin{figure}[t]
  \centering
  \includegraphics[width=\columnwidth]{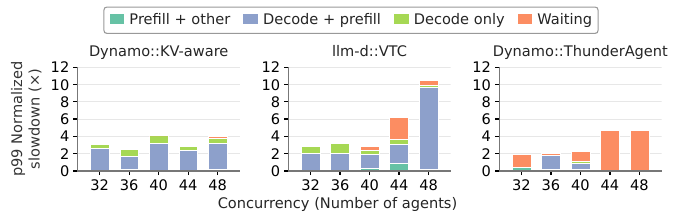}
  \vspace{-0.75cm}
\caption{Breakdown of p99 normalized slowdown across TraceLab routing policies into Prefill + other (prompt processing and residual time), Decode + prefill (decoding during concurrent prefill), Decode only, and Waiting (queuing). Admission control trades interference for router waiting time.}

\label{fig:router-rtx}
\Description{}
\end{figure}
Figure~\ref{fig:router-rtx} decomposes p99 normalized slowdown across these routing policies.
Dynamo::KV-aware preserves cache affinity but lacks admission control, allowing memory pressure and prefix recomputation to slow co-running requests.
llm-d::VTC rotates admission to balance token-level service, but KV cache eviction and recomputation increase prefill interference at high concurrency.
In contrast, Dynamo::ThunderAgent limits working sets, reducing execution interference but shifting unfairness into admission waiting.
At high concurrency, its most delayed agents, particularly short-running ones, mostly wait for GPU admission rather than execution.
Thus, admission policies can reduce worker interference while disproportionately delaying agents that have yet to be admitted.

\begin{figure}[t]
  \centering
  \includegraphics[
        width=0.9\columnwidth, trim=8pt 0pt 0pt 0pt, clip]{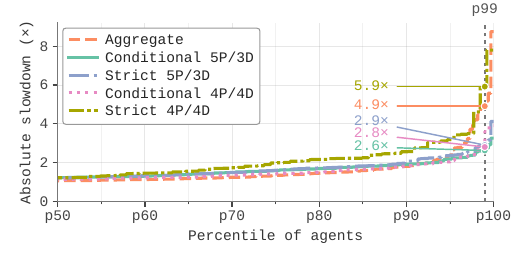}
        \vspace{-0.75cm}
  \caption{Agent slowdown under aggregated serving and strict or conditional PD disaggregation on eight MI300X.}

  \label{fig:agent-pd}
  \Description{}
\end{figure}
We next examine how PD disaggregation policies affect fairness.
Figure~\ref{fig:agent-pd} compares aggregated serving against strict and conditional PD disaggregation~\cite{dynamo} on 8$\times$ AMD MI300X GPUs\footnote{To study PD disaggregation, we use an MI300X server with AMD's Infinity Fabric (xGMI) interconnect~\cite{kolla2025xgmi}, similar to NVLink, which the RTX GPU does not support.}, using Qwen with TP=1, EP=1, and FCFS scheduling.
Strict disaggregation sends every prefill to prefill workers, while conditional disaggregation offloads a prefill only when its uncached prompt length exceeds a threshold and otherwise runs it locally on the decode worker.
We replay Azure traces with 20 concurrent agents for 60 minutes, evaluating 5P/3D and 4P/4D allocations.
With 5P/3D, strict and conditional disaggregation reduce p99 slowdown from 4.9$\times$ to 2.9$\times$ and 2.6$\times$, respectively.
However, the benefit depends on capacity allocation: under 4P/4D, conditional disaggregation achieves 2.8$\times$, whereas strict disaggregation worsens slowdown to 5.9$\times$ due to prefill queuing.
Even with effective disaggregation, residual interference persists among concurrent prefills or when conditional placement overlaps prefill and decode.
Thus, eliminating cross-phase interference alone does not guarantee fairness; the placement policy must also balance phase-specific demand.
\begin{center}
\begin{tcolorbox}[
    width=\columnwidth,
    boxsep=1pt,
    left=2pt,
    right=2pt,
    top=4pt,
    bottom=4pt,
    halign=center,
    fontupper=\small
]
\emph{\textbf{Takeaway \#3:} Routing policies can shift unfairness from GPU contention to admission waiting, while PD disaggregation can introduce new queuing bottlenecks if capacity is poorly allocated.}
\end{tcolorbox}
\end{center}

\section{Optimizing Fairness using \sys{}}
\label{sec:optimization}



We now present \sys{}, a fairness-aware serving system designed to address the sources of unfairness identified in Section~\ref{sec:attributing_unfrirness}.

\subsection{Optimization Objective}
As demonstrated in Takeaway~\#2, throughput-oriented optimizations such as KV cache retention can improve efficiency while disproportionately slowing some agents.
In this paper, \textit{for a given LLM and GPU architecture}, we focus on optimizing how agents share serving resources through routing, admission control, and runtime scheduling.
To balance fairness and efficiency, \textbf{we formulate the optimization objective as minimizing p99 agent-level normalized slowdown subject to a throughput constraint}:
\begin{equation}
\label{eq:objective}
\min_{\pi} \; p_{99}\!\left(\{U_r\}\right)
\quad \text{s.t.} \quad
X(\pi) \ge (1-\epsilon)\, X_{\text{baseline}},
\end{equation}
where $\pi$ is the serving policy, $U_r$ is the normalized slowdown of the task rooted at agent $r$, $X(\pi)$ is the throughput, and $X_{\text{baseline}}$ is the throughput of a performance-optimized baseline.
The parameter $\epsilon$ is the desired throughput tolerance, although \sys{} aims to minimize throughput degradation rather than guarantee a strict bound.

\subsection{Design Overview}
\label{sec:design-overview}

Figure~\ref{fig:duru-overview} illustrates the design of \sys{}, which coordinates fairness tracking, scheduling, admission control, and autoscaling.
As demonstrated in Takeaway~\#3, reducing resource contention at one serving layer can shift unfairness into queuing delays at another.
This motivates \sys{}'s cross-layer design, which uses agent-level slowdown as a common control signal to coordinate router and runtime decisions rather than optimizing each layer independently.
The design comprises four key components:

\begin{figure}[t]
  \centering
  \includegraphics[width=\columnwidth]{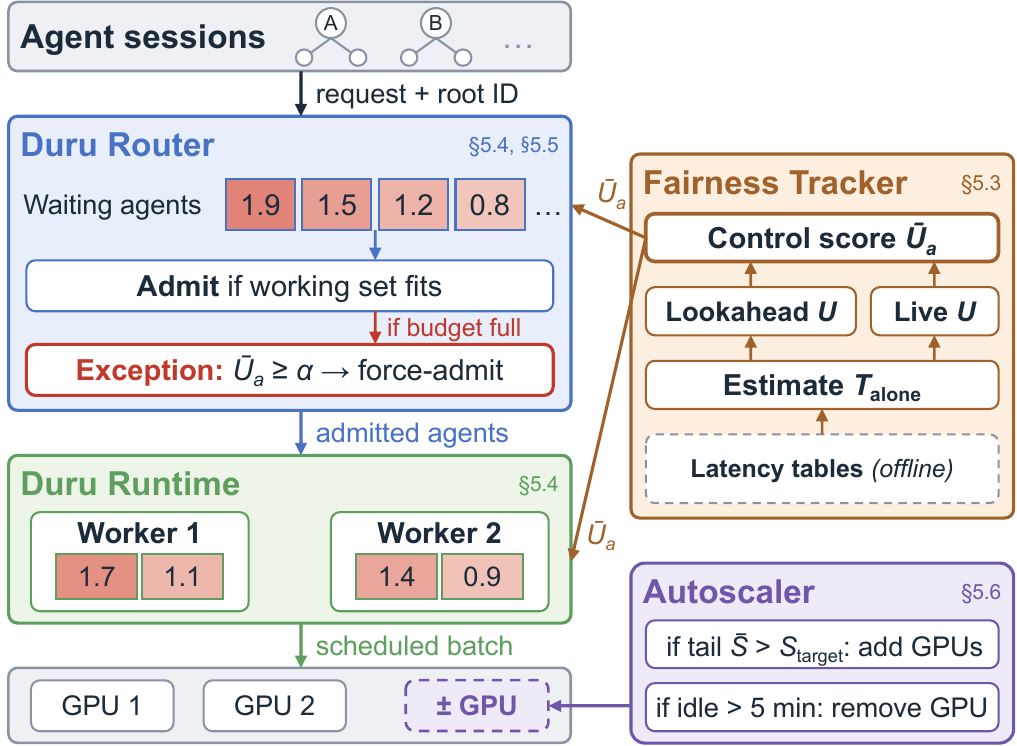}
  \vspace{-0.65cm}
  \caption{Overview of \sys{}. The fairness tracker computes each agent's control
  score $\bar{U}_a$ (\S\ref{sec:control-signal}). The router and workers order waiting
  agents and requests by $\bar{U}_a$ (\S\ref{sec:ordering}), and the router force-admits
  starved agents (\S\ref{sec:exceptions}). The autoscaler adds GPUs when the p99 of 
  $\bar{S}_a$ exceeds $S_{\text{target}}$ (\S\ref{sec:autoscaling}).}
          \vspace{-0.5cm}

    \Description{}
  \label{fig:duru-overview}
\end{figure}
\begin{itemize}[leftmargin=*]
    \item \textbf{Online Fairness Tracker.}
    \sys{} continuously estimates each agent's normalized slowdown using observed waiting time, estimated $\Talone$, and completed inference work (\S\ref{sec:control-signal}).
    The score $\bar{U}_a$ tracks each agent’s deficit from its fair share.
    \item \textbf{Slowdown-Aware Scheduling.}
    The router and runtime prioritize agents with higher $\bar{U}_a$, coordinating scheduling across both layers (\S\ref{sec:ordering}).
    These decisions prioritize delayed agents while respecting KV cache capacity to preserve throughput.
    \item \textbf{Starvation-Aware Admission.}
    Under memory pressure, strict admission limits can indefinitely delay agents despite slowdown-aware ordering.
    \sys{} introduces admission exceptions when an agent's score exceeds a threshold, temporarily relaxing the router's working-set budget to guarantee progress (\S\ref{sec:exceptions}).
    \item \textbf{SLO-Driven Autoscaling.}
    Scheduling and admission redistribute resources to improve relative fairness but cannot eliminate delays caused by insufficient capacity.
    To meet SLO, \sys{} adds GPUs when p99 live slowdown exceeds $S_{\text{target}}$ (\S\ref{sec:autoscaling}).
\end{itemize}

Together, these components reduce disproportionate delays through coordinated scheduling and admission, while autoscaling addresses capacity shortfalls that scheduling alone cannot resolve.




\subsection{Control Signal}
\label{sec:control-signal}

\sys{} translates the agent-level normalized slowdown metric (\S\ref{sec:metric}) into an online control signal for fairness-aware scheduling.
Its fairness tracker maintains a score $\bar{U}_a$ for each agent, which the router and workers use to prioritize execution.
A higher score identifies an agent further behind its fair share, allowing \sys{} to prioritize disproportionately delayed agents.
Unlike metrics computed after job completion, this signal is continuously updated using observed execution progress, without requiring knowledge of future requests.

For agents that have begun execution, $\bar{U}_a$ is their live normalized slowdown (Definition~\ref{def:live-slowdown}), computed from accumulated waiting and execution time relative to estimated standalone work (\S\ref{sec:talone}).
Tracking this signal incurs little overhead: existing serving infrastructure already maintains request-level metadata, such as timestamps and token counts~\cite{kwon2023vllm,zheng2024sglang}, while \sys{} only requires a few additional per-agent counters for accumulated service entitlement and estimated completed work.
In our measurements, this bookkeeping has a negligible 0.3--0.8\% time overhead.
These counters are updated incrementally, requiring constant-size metadata per agent rather than storing its execution history.
For multi-agent jobs, the tracker aggregates the root agent and its sub-agents into a single score.

For newly arriving agents, live slowdown is initially undefined because no inference work has been completed.
To initialize their scores, \sys{} uses \emph{lookahead} $U$, which includes the estimated standalone cost of the request's next five engine iterations in the denominator.
This estimate uses the same latency models (\S\ref{sec:talone}), enabling newly arriving agents to participate in slowdown-based scheduling.

\begin{algorithm}[t]
\caption{\sys{}: End-to-End Fairness-Aware Serving}
\label{alg:duru}
\small
\begin{algorithmic}[1]
\Require 
\Statex $Q,A$: worker and router waiting queues
\Statex $\bar U_a,\bar S_a$: fairness and SLO signals
\Statex $B_j,W_j,w_a$: KV budget, usage, and agent demand
\Statex $\mathcal G$: active GPU workers
\Statex $\alpha,S_{\mathrm{target}}$: admission and SLO thresholds

\Function{ScheduleWaiting}{$Q$} \Comment{\textcolor{blue}{runtime worker}}
  \For{$q \in Q$ in decreasing $\bar U$}
    \If{KV allocation for $q$ succeeds}
      \State schedule $q$
    \Else
      \State defer $q$
    \EndIf
  \EndFor
\EndFunction

\Function{AdmitCandidates}{$A$} \Comment{\textcolor{blue}{router}}
  \For{$a \in A$ in decreasing $\bar U$}
    \If{$\exists j: W_j+w_a \le B_j$}
      \State admit $a$ to $j$
    \ElsIf{$a$ is pending and $\bar U_a \ge \alpha$}
      \State force-admit $a$
    \Else
      \State defer $a$
    \EndIf
  \EndFor
  \For{each worker $j$ with $W_j>B_j$}
    \State pause smallest-$w_a$ tool-phase agents
    \Statex \hspace{\algorithmicindent} until $W_j \le B_j$
  \EndFor
\EndFunction

\Function{Autoscale}{$\mathcal G$} \Comment{\textcolor{blue}{autoscaler}}
  \If{$p_{99}(\{\bar S_a\}) \ge S_{\mathrm{target}}$}
      \State add one GPU worker
  \ElsIf{$\exists j \in \mathcal G$: idle $\ge 300$\,s}
      \State remove worker $j$
  \EndIf
\EndFunction

\end{algorithmic}
\end{algorithm}

\subsection{Slowdown-Based Ordering}
\label{sec:ordering}

\sys{} prioritizes agents based on their accumulated unfairness, using the control score $\bar{U}_a$ to determine scheduling order.
Both the router and runtime consider waiting candidates in decreasing $\bar{U}_a$, breaking ties by arrival order.
This ordering dynamically adapts to each agent's execution progress: an agent experiencing waiting accumulates a higher score, increasing its priority over agents receiving more favorable service.
This ensures that scheduling decisions account for accumulated agent-level delays rather than relying solely on immediate execution efficiency.

To preserve throughput, \sys{} combines slowdown-based ordering with capacity-aware scheduling at both layers.
At the router, agents are admitted subject to a working-set budget and resume hysteresis, while workers schedule requests according to KV cache capacity and scheduling budgets.
At either layer, candidates that cannot fit are skipped, allowing lower-priority candidates to backfill unused capacity.
This avoids head-of-line blocking and maintains resource utilization without abandoning fairness priorities.
Skipped candidates are reconsidered in subsequent scheduling rounds as their control scores evolve.
However, when memory pressure persistently prevents admission, ordering alone cannot ensure timely progress, motivating the admission exceptions discussed next.




\subsection{Admission Exceptions}
\label{sec:exceptions}
Slowdown-based ordering prioritizes delayed agents, but it cannot guarantee their progress when the router's working-set budget is exhausted.
In particular, long-running agents with retained KV caches can occupy the memory, preventing other agents from being admitted even as their slowdown increases.
To address this, \sys{} introduces an admission-exception threshold $\alpha$, which determines when an agent's accumulated unfairness warrants overriding admission limits.
When a waiting agent cannot be admitted under the working-set budget and its control score satisfies $\bar{U}_a \ge \alpha$, \sys{} force-admits it to an eligible worker.
This relaxes the working-set budget to allow delayed agents to make progress.
As admitted agents execute and others are paused, the system returns toward its normal working-set budget, limiting sustained memory pressure.

The threshold $\alpha$ provides a mechanism to balance tail fairness against the throughput benefits of KV cache reuse.
A smaller $\alpha$ triggers earlier and more frequent interventions, reducing prolonged waiting but potentially increasing KV cache eviction, recomputation, and scheduling overhead.
Conversely, a larger $\alpha$ favors cache retention and throughput by admitting fewer exceptions, at the risk of longer delays for some agents.
Overly aggressive admission can worsen both throughput and fairness if the resulting cache churn outweighs the benefits of reducing waiting.
Thus, $\alpha$ tunes fairness intervention against throughput degradation.
A natural choice is $\alpha = 1$, since $U = 1$ corresponds to receiving the PS fair share (\S\ref{def:ps-fairness}).



\subsection{SLO-Driven Autoscaling}
\label{sec:autoscaling}
Optimizing normalized slowdown improves fairness across agents but does not guarantee that agents meet their performance SLOs, particularly when system load exceeds available capacity.
Traditional LLM serving SLOs, such as Time-To-First-Token (TTFT) and Time-Between-Tokens (TBT), primarily target human-facing interactions and do not adequately capture progress in autonomous agentic workloads.
Instead, we adopt a slowdown-based SLO, following prior studies that express latency targets relative to isolated execution time~\cite{li2023alpaserve,stojkovic2025dynamollm}.
This formulation suits heterogeneous agentic tasks because it scales each task's latency target by its standalone execution time.
We therefore define SLO attainment at the agent-task level as completing inference within $S_{\text{target}}$ times its standalone execution time, using $S_{\text{target}}=5$ in our evaluation.
Unlike the SLO, normalized slowdown $U$ measures whether an agent receives its fair share of available capacity under the prevailing system load.
Thus, fairness guides resource sharing at a given capacity, whereas the SLO guides capacity scaling to meet a specified slowdown target.

To improve SLO attainment under varying workloads, \sys{} introduces SLO-driven autoscaling that adjusts GPU capacity based on observed agent performance rather than resource utilization alone.
The autoscaler monitors active agents to identify potential SLO violations and adds a GPU worker when the slowdown is close to exceeding the SLO target.
This targets capacity-induced delays that scheduling and admission control alone cannot eliminate.
To avoid oscillations and unnecessary provisioning, \sys{} scales out incrementally, and scales in only after a worker has remained idle for five minutes with no requests waiting in the router.
Algorithm~\ref{alg:duru} summarizes \sys{}'s end-to-end workflow, integrating slowdown-aware runtime scheduling, starvation-aware admission control, and SLO-driven autoscaling.




\section{Evaluation Methodology}
\label{sec:methodology}

Our evaluation seeks to answer four key questions.
\textbf{(Q1)} Does \sys{} improve p99 fairness without significantly sacrificing throughput?
\textbf{(Q2)} How do router- and runtime-level optimizations contribute to end-to-end fairness?
\textbf{(Q3)} How does \sys{}'s fairness--throughput trade-off scale with increasing GPU fleet size?
\textbf{(Q4)} Can \sys{} meet agent-level SLOs through autoscaling under dynamic workloads?

\begin{figure*}[t]
  \centering
  \includegraphics[width=\textwidth]{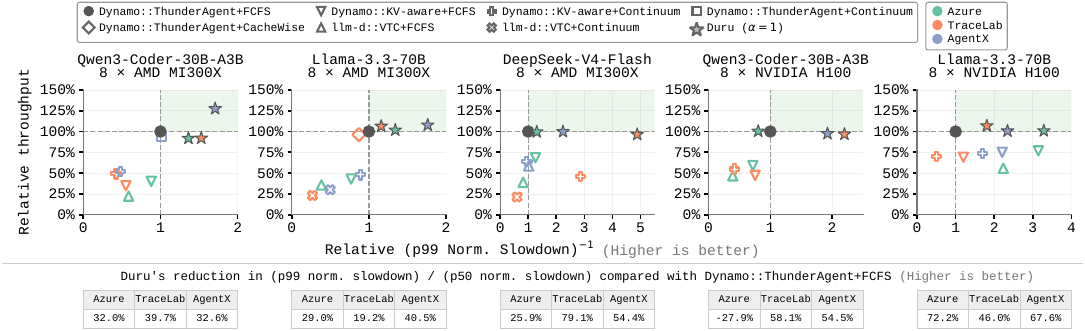}
  \vspace{-0.5cm}
    \caption{Throughput versus inverse p99 normalized slowdown, relative to Dynamo::ThunderAgent+FCFS. \textbf{Key takeaway:} \sys{} improves p99 fairness while largely preserving throughput across diverse models, workloads, and GPU architectures.}
    \vspace{-0.25cm}
  \label{fig:matrix-pareto}
\Description{}
\end{figure*}

\begin{table}[t]
\centering
\caption{Evaluation setup.}
\vspace{-0.35cm}
\label{tab:exp-matrix}
\footnotesize
\setlength{\tabcolsep}{2pt}
\renewcommand{\arraystretch}{1.05}
\begin{tabularx}{0.48\textwidth}{
    @{}
    >{\raggedright\arraybackslash}p{1.35cm}
    >{\raggedright\arraybackslash}p{2.65cm}
    >{\raggedright\arraybackslash}X
    @{}
}
\toprule
 & \textbf{Name} & \textbf{Description} \\
\midrule
\textbf{Models}
  & Qwen3-Coder-30B-A3B~\cite{qwen2025qwen3}
  & MoE, 30.5B total, 3.3B active  \\ 
  & Llama-3.3-70B~\cite{meta2024llama33}
  & Dense, 70B parameters  \\ 
  & DeepSeek-V4-Flash~\cite{deepseek2026v4}
  & MoE, 284B total, 13B active  \\ 
\midrule
\textbf{Workloads}
  & Azure~\cite{liu2026azure}
  & Copilot coding-agent sessions \\
  & TraceLab~\cite{zhu2026tracelab}
  & Claude Code and Codex sessions \\
  & AgentX~\cite{semianalysis2026agentx}
  & SemiAnalysis agentic workload traces \\
\midrule
\textbf{Platforms}
  & NVIDIA H100 SXM
  & 8 GPUs, 80\,GB HBM3 each \\
  & AMD MI300X
  & 8 GPUs, 192\,GB HBM3 each \\
  & DynoSim~\cite{ding2026dynosim}
  & NVIDIA Dynamo simulator~\cite{dynamo} \\
\midrule
\textbf{Software}
  & LMCache 0.5.4~\cite{liu2025lmcache}
  & CPU DRAM offload, 128GiB per worker\\
  & vLLM 0.28.0~\cite{kwon2023vllm} 
  & LLM serving engine\\
\bottomrule
\end{tabularx}
\end{table}
To answer these questions, we evaluate \sys{} across diverse LLM architectures, real-world agentic workloads, and GPU platforms from multiple vendors, as summarized in Table~\ref{tab:exp-matrix}.
We use three models spanning sizes and computational characteristics: Qwen3-Coder-30B-A3B (30B parameters, MoE), Llama-3.3-70B (70B parameters, dense), and DeepSeek-V4-Flash (284B parameters, MoE with compressed attention).
This diversity allows us to evaluate fairness under different compute, memory, and KV cache demands.
For workloads, we use three real-world production agentic execution traces: Azure GitHub Copilot~\cite{liu2026azure}, TraceLab~\cite{zhu2026tracelab}, and AgentX~\cite{semianalysis2026agentx}.
These traces capture complex, multi-turn agentic workflows, including both single-agent and multi-agent execution, representative of contemporary AI serving workloads.
Our main evaluation uses NVIDIA H100 and AMD MI300X GPUs.\footnote{Our characterization (\S\ref{sec:attributing_unfrirness}) additionally uses NVIDIA H200 and RTX PRO 6000 Blackwell GPUs to study contention and serving behavior across different hardware.}
Finally, we use DynoSim~\cite{ding2026dynosim} to evaluate scalability across clusters of 8 to 256 GPUs and study autoscaling under production arrival patterns.

\noindent\textbf{Baselines.}
We compare \sys{} against the state-of-the-art scheduling and routing policies at both layers of the serving stack.
At the runtime layer, we evaluate FCFS scheduling, Continuum~\cite{li2026continuum}, which retains KV caches across tool calls using time-to-live estimates, and CacheWise~\cite{tiwari2026cachewise}, which prioritizes requests with low KV cache demand and evicts blocks based on future reuse.
We use the oracle variant of CacheWise.
At the routing layer, we evaluate NVIDIA Dynamo's KV-aware routing~\cite{dynamo}, which prioritizes KV cache locality; llm-d::VTC~\cite{llmd-wfs}, which incorporates token-level fairness into admission control~\cite{sheng2024vtc}; and ThunderAgent~\cite{kang2026thunderagent}, which uses program-level working-set accounting to control admission.
NVIDIA Dynamo and llm-d are production-grade serving frameworks deployed in data centers today.
By evaluating their routing and admission policies alongside research baselines, we assess \sys{} against approaches representative of serving infrastructure.

\noindent\textbf{Memory Pressure.}
To evaluate fairness under resource contention, we configure concurrency to target an aggregate working-set demand of approximately $1.3\times$ the available GPU KV cache capacity.
Due to variations in agent context lengths, actual demand ranges from $1.2\times$ to $1.5\times$ capacity across workloads.
For each model--workload combination, we maintain identical concurrency across all policies to ensure a fair comparison.
Host-memory KV cache offloading using LMCache~\cite{liu2025lmcache} is enabled for all main comparisons.

\noindent\textbf{Scalability and autoscaling.}
We use DynoSim~\cite{ding2026dynosim} to simulate clusters of 8--256 NVIDIA H100 GPUs, each running one TP1 Qwen (FP8) worker.
We replay Azure and AgentX traces with inter-request gaps capped at 30\,s.
For autoscaling, we replay 48 hours of Azure traffic with UTC arrivals and causal request dependencies.
To minimize prefill--decode interference, we use disaggregated pools, with 32 GPUs per phase, scaling between 8 and 128 GPUs in total.
All policies share the slowdown-driven autoscaler (\S\ref{sec:autoscaling}) with $S_{\mathrm{target}}=5$~\cite{stojkovic2025dynamollm}, adding one GPU with a 30\,s cooldown and removing workers only after 300\,s of idleness with no queued requests for that phase.
We model worker startup delays of 480\,s (cold) and 60\,s (warm), with released nodes remaining warm for 600\,s; workers serve requests after startup completes.
Host-memory KV cache offloading is disabled in simulations due to DynoSim limitations.
\section{Evaluation}
\label{sec:evaluation}

\noindent\underline{\textbf{Q1: Throughput--fairness trade-off.}}
Figure~\ref{fig:matrix-pareto} compares the fairness--throughput trade-off of \sys{} against different routing and runtime policies across three models, three workloads, and AMD MI300X and NVIDIA H100 GPUs, using $\alpha=1$.
Because throughput and slowdown vary across models and workloads, we normalize both metrics against Dynamo::ThunderAgent with FCFS for each configuration.
The X-axis shows the inverse of relative p99 normalized slowdown ($U_{\mathrm{baseline}}/U_{\mathrm{policy}}$); higher values indicate better fairness, while the Y-axis shows relative throughput, where higher is better.
The dashed lines indicate baseline performance, and the shaded upper-right region represents improvements in both fairness and throughput.
\sys{} achieves more favorable operating points than most competing configurations, pushing the fairness--throughput trade-off outward.
Across the 15 model–workload–hardware settings evaluated on MI300X and H100, Duru reduces p99 normalized slowdown and the p99/p50 ratio by geometric means of 44.4\% and 46.7\%, respectively, while maintaining 91.8–127.4\% of baseline throughput.
In 14 of these settings, Duru reduces p99 normalized slowdown by 13.8–79.5\% and the p99/p50 ratio by 19.2–79.1\%.
In contrast, other state-of-the-art policies achieve lower throughput, worse fairness, or both.

The trade-off between fairness and throughput varies across workloads and models.
On MI300X, \sys{} improves fairness for Qwen on Azure and TraceLab with modest throughput reductions, while AgentX benefits in both dimensions.
For Llama, fairness and throughput improve across all workloads on both platforms.
DeepSeek also achieves better fairness across all workloads on MI300X while largely preserving throughput, with particularly pronounced fairness gains on TraceLab and AgentX.
On H100, Qwen achieves substantial fairness gains on TraceLab and AgentX with small throughput reductions, but exhibits worse fairness on Azure with essentially unchanged throughput.

Azure’s weaker fairness stems from long-prefill interference.
Azure has 2.3$\times$ as many appended input tokens per output token as TraceLab.
In the Qwen H100 run, tail requests queue behind long prefills and decode alongside them, increasing slowdown.
Although \sys{} reduces queuing through slowdown-aware scheduling and admission, it cannot eliminate the prefill interference with concurrent decoding.
Consequently, fairness improvements are strongest when waiting dominates the tail, whereas prefill interference can limit the benefit of scheduling alone.
These results highlight that improving agent-level fairness requires addressing both unfair resource allocation and the sources of execution interference.

\begin{figure}[t]
  \centering
  \includegraphics[width=\columnwidth]{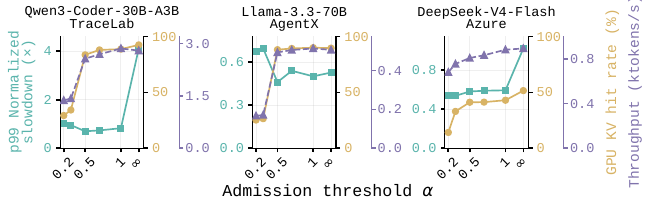}
  \vspace{-0.75cm}
  \caption{Sensitivity analysis: effect of admission-exception threshold $\alpha$ on fairness, throughput, and GPU KV hit rate.}
  \vspace{-0.2cm}
  \label{fig:alpha-variant}
\Description{}
\end{figure}
\noindent\textbf{Sensitivity to parameter $\alpha$.}
While Figure~\ref{fig:matrix-pareto} uses $\alpha=1$, Figure~\ref{fig:alpha-variant} examines how different thresholds affect fairness.
The primary Y-axis shows p99 normalized slowdown, while the secondary Y-axis shows throughput and KV cache hit rate to explain the underlying performance changes.
At $\alpha=\infty$, admission exceptions are disabled, resulting in high tail slowdown despite high cache hit rates.
Reducing $\alpha$ generally improves fairness by admitting delayed agents sooner, but increases competition for GPU memory.
When $\alpha$ becomes too small, KV cache hit rates can drop sharply, increasing recomputation, dropping throughput, and potentially worsening fairness.
\emph{Importantly, $\alpha=1$ achieves consistently strong fairness across the evaluated workloads, suggesting that a fixed threshold suffices without sophisticated dynamic tuning.}

\begin{figure}[t]
  \centering
  \includegraphics[width=\columnwidth]{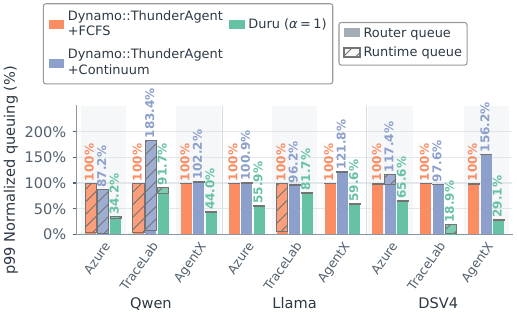}
  \vspace{-0.75cm}
  \caption{p99 normalized queuing delay relative to ThunderAgent+FCFS (100\%) for each model–workload pair.}
  \vspace{-0.25cm}
  \label{fig:matrix-decomp}
\Description{}
\end{figure}
\noindent\textbf{Attributing \sys{}'s fairness improvements.}
To understand why \sys{} improves fairness, Figure~\ref{fig:matrix-decomp} decomposes the p99 normalized queuing delay, $p_{99}(\{T_i^{\mathrm{queue}}/T_i^{\mathrm{alone}}\})$, into router and runtime waiting on AMD MI300X GPUs, normalized to Dynamo::ThunderAgent with FCFS.
Across workloads, the dominant source of waiting varies: admission limits delay at the router, while resource contention delays admitted requests at the runtime.
Importantly, reductions in p99 queuing delay correspond to improved p99 fairness (Figure~\ref{fig:matrix-pareto}), but not proportionally, as the tail agents in these two distributions may differ.
By prioritizing agents based on their accumulated slowdown across both layers and introducing admission exceptions, \sys{} mitigates disproportionate waiting rather than merely shifting delays between queues.
\textit{This underscores the importance of cross-layer fairness mechanisms over single-layer optimizations.}

\begin{figure}[t]
  \centering
  \includegraphics[width=\columnwidth]{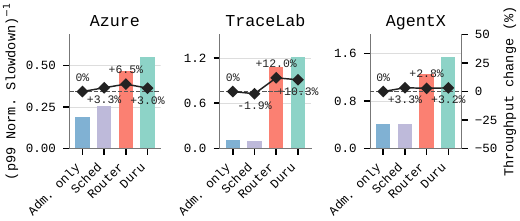}
  \vspace{-0.75cm}
  \caption{Duru ablation study on Qwen across three traces: inverse p99 normalized slowdown and throughput change.}
  \vspace{-0.5cm}
  \label{fig:ablation}
\Description{}
\end{figure}

\noindent\underline{\textbf{Q2: Ablation study of cross-layer fairness.}}
To quantify the contribution of each layer's fairness mechanisms, Figure~\ref{fig:ablation} compares four configurations on Qwen: admission control without slowdown ordering and admission exceptions (\emph{Adm. only}), runtime-only fairness (\emph{Sched}), router-only fairness (\emph{Router}), and their combination (\sys{}).
The bars show the inverse of p99 normalized slowdown, and markers indicate throughput relative to \emph{Adm. only}.

The results show that each mechanism's effectiveness depends on workload characteristics.
On Azure, both runtime and router optimizations contribute to fairness, with their combination achieving the best result.
Azure's longer prefills increase runtime contention, making slowdown-aware scheduling important.
In contrast, TraceLab and AgentX contain longer-running sessions whose retained KV caches occupy the available working-set budget, concentrating waiting at the router.
Runtime scheduling alone provides little benefit in these workloads.
However, router-level admission exceptions introduce additional runtime contention, which slowdown-aware scheduling helps mitigate.
\emph{These results demonstrate that coordinating router admission and runtime scheduling improves fairness by addressing their interacting effects on resource contention.}

\begin{figure}[t]
  \centering
  \includegraphics[width=\columnwidth]{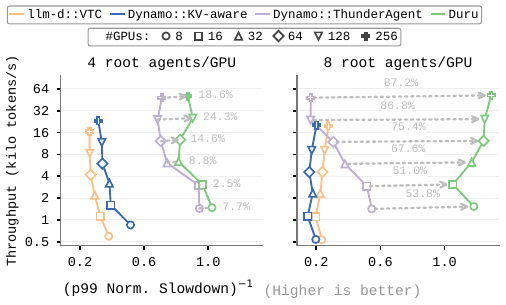}
  \vspace{-0.75cm}
    \caption{Throughput–fairness trade-off of AgentX. Fleet scales from 8 to 256 GPUs, with 4 or 8 root agents per GPU.}
\label{fig:duru-scaling}
\Description{}
\end{figure}
\noindent\underline{\textbf{Q3: Scalability analysis.}}
Figure~\ref{fig:duru-scaling} compares the inverse of p99 normalized slowdown (higher is better) against throughput as GPU fleets scale from 8 to 256 GPUs, with 4 or 8 root agents per GPU.
\sys{} maintains throughput comparable to the strongest baseline, Dynamo::ThunderAgent, while its fairness advantage generally grows with fleet size.
This improvement is particularly pronounced under higher concurrency, where \sys{} reduces p99 normalized slowdown by up to 87.2\%.
Interestingly, even with constant per-GPU concurrency, fairness can deteriorate as the fleet grows.
This occurs because stateful agents tend to remain on workers holding their KV caches, creating persistent load imbalance across workers.
Larger fleets increase the likelihood of overloaded workers and disproportionately delayed agents, amplifying tail slowdown.
By coordinating fairness-aware admission and scheduling, \sys{} mitigates these delays while preserving KV cache reuse.
\emph{This shows that scaling GPU capacity alone does not eliminate unfairness; \sys{} preserves its fairness--throughput benefits as the fleet grows.}

\begin{figure}[t]
  \centering
  \includegraphics[width=\columnwidth]{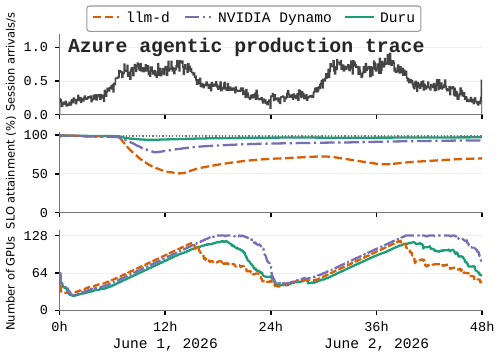}
  \vspace{-0.75cm}
  \caption{Azure agentic trace replay showing \sys{} achieves improved SLO attainment at lower GPU fleet size.}
          \vspace{-0.5cm}

  \label{fig:autoscale}
\Description{}
\end{figure}
\noindent \underline{\textbf{Q4: Effectiveness of autoscaling in meeting SLO.}}
Figure~\ref{fig:autoscale} evaluates llm-d, NVIDIA Dynamo, and \sys{} replaying 48 hours of Azure GitHub Copilot production traces~\cite{liu2026azure}.
The figure shows session arrival (top), SLO attainment (middle), and \#GPUs (bottom).
We use the slowdown-based SLO from \S\ref{sec:autoscaling} with $S_{\mathrm{target}}=5$.\footnote{\sys{} also outperforms the baselines at other thresholds, though by smaller margins as the target is relaxed; we report only $S_{\mathrm{target}}=5$ for space limitation.}
All policies use the same autoscaler, allowing us to evaluate fairness-aware scheduling and admission on SLO attainment.

\sys{} achieves the highest SLO attainment of 97.4\%, compared with 93.4\% for NVIDIA Dynamo and 70.2\% for llm-d, with nearly identical throughput.
\sys{} achieves this improvement using 76 GPUs on average, compared with 89 GPUs for NVIDIA Dynamo.
This benefit arises because fairness-aware scheduling changes which agents are admitted and co-scheduled on GPUs, reducing disproportionate waiting without added capacity.
When arrivals spike, startup delays temporarily constrain capacity, causing attainment drops; \sys{} prioritizes delayed agents and maintains near-constant SLO attainment.
Conversely, GPU scale-in lags declining demand because stateful agents retain contexts on workers until they become idle.
\emph{These results show that \sys{} effectively combines fairness optimization with autoscaling to achieve high SLO attainment and efficient GPU utilization under production workloads.}

\section{Related Work}
\label{sec:related}



\noindent\textbf{Fairness in resource scheduling.}
Prior work on shared-hardware fairness uses application slowdown relative to isolated execution to characterize interference and guide resource allocation~\cite{mutlu2007stfm,mutlu2008parbs,ebrahimi2010fst,subramanian2015asm}.
Meanwhile, OS schedulers~\cite{linux2007cfs,linux2024eevdf} and network schedulers~\cite{parekh1993gps,demers1989wfq,shreedhar1995drr} approximate PS, while dominant resource fairness (DRF)~\cite{ghodsi2011drf} generalizes fair allocation to multiple resources.
Building on these principles, \sys{} adopts slowdown relative to isolated execution and uses PS as its fairness reference (Definition~\ref{def:ps-fairness}), extending fairness measurement and optimization to agentic LLM serving.

\noindent\textbf{Fairness in LLM serving.}
VTC~\cite{sheng2024vtc} and DLPM~\cite{cao2025dlpm} balance weighted token service; DLPM additionally exploits prefix locality.
FairInference~\cite{bali2026fairinference} provides per-token latency isolation; FairCache~\cite{bai2026faircache} accounts for the compute, memory, and I/O costs in context caching.
Token-based metrics miss execution costs (\S\ref{sec:hardware-layer}); request-level metrics miss cumulative latency (Figure~\ref{fig:matrix-decomp}).

\noindent\textbf{Agent-aware serving.}
Recent systems optimize KV cache retention~\cite{li2026continuum,xia2026idleness,tiwari2026cachewise}, prefix sharing~\cite{pan2025kvflow,bian2026tokencake,pan2026scalesim}, and program-aware scheduling~\cite{kang2026thunderagent,luo2026agentix,liu2026askthetool}.
SMetric~\cite{wang2026smetric} considers session-level service balance, while SAGA~\cite{guo2026saga} introduces workflow-level fairness using predicted remaining work and task deadlines.
Unlike prediction-based scheduling, \sys{} tracks fairness through observed slowdown and requires no knowledge of future.
It further coordinates admission and scheduling to address cumulative delays.

\section{Conclusion}
\label{sec:conclusion}
Agent-level fairness has received limited attention despite the adoption of shared LLM serving infrastructure.
In this work, we defined agent-level fairness using normalized slowdown relative to standalone execution, characterized its sources, and designed \sys{} to optimize it.
Our analysis revealed hardware contention, KV cache retention, and routing decisions caused disproportionate slowdowns, with queuing and starvation often dominating tail performance.
To address these challenges, \sys{} coordinated slowdown-aware routing and runtime scheduling, introduced admission exceptions to mitigate starvation, and incorporated autoscaling to improve SLO attainment under varying workloads.
\sys{} improved p99 fairness by a geometric mean of 44.4\% (up to 79.5\%) while achieving 91.8--127.4\% of baseline throughput.
It also achieved 97.4\% SLO attainment under production arrivals.
\textit{These results demonstrate that coordinated, cross-layer fairness control substantially reduces agent-level unfairness while preserving serving efficiency.}
\section{Acknowledgement}
\label{sec:acknowledgement}
Generative AI tools were used for manuscript editing, formatting, and code development.
The authors take full responsibility for the ideas and the accuracy of the results.
The authors acknowledge insightful discussions with Vikram Sharma Mailthody, Krisztian Flautner, Mosharaf Chowdhury, and Rakesh Kumar.
This work was supported in part by the AMD AI and HPC Cluster Program, AMD's Funding Academic Research program, funding from Gimlet Labs, and DARPA ML2P Award HR0011-26-9-E190.
This work used the Delta advanced computing resource at the National Center for Supercomputing Applications (NCSA) through the National Science Foundation’s Advanced Cyber infrastructure Coordination Ecosystem: Services and Support (ACCESS) program under allocation CIS261278.


\bibliographystyle{ACM-Reference-Format}
\bibliography{refs}

\end{document}